\documentclass{article}

\usepackage[utf8]{inputenc}
\usepackage[margin=1in]{geometry}
\usepackage[titletoc,title]{appendix}
\usepackage{booktabs}
\usepackage{caption}
\usepackage{subcaption}
\usepackage{amsmath,amsfonts,amssymb,mathtools}

\usepackage{graphicx,float}
\usepackage[affil-it]{authblk}

\usepackage[ruled,vlined]{algorithm2e}
\usepackage{algorithmic}
\usepackage[utf8]{inputenc}

\usepackage{biblatex}
\title{Subduction zone fault slip from seismic noise and GPS data}
\author{José Augusto Proença Maia Devienne}
\affil{Big Data to Earth Scientists}
\date{November, 2020}

\begin{document}

\maketitle

\section{Introduction}

In Geosciences a class of phenomena that is widely studied given its real impact on human life are the tectonic faults slip. These landslides have different ways to manifest, ranging from aseismic events of slow displacement (slow slips) to ordinary earthquakes. An example of continuous slow slip event was identified in Cascadia, near the island of Vancouver [1]. This slow slip event is associated with a tectonic movements, when the overriding North America plate lurches southwesterly over the subducting Juan de Fuca plate. This region is located down-dip the seismogenic rupture zone, which has not been activated since 1700s but has been cyclically loaded by the slow slip movement. This fact requires some attention, since slow slip events have already been reported in literature as possible triggering factors for earthquakes [2]. Nonetheless, the physical models to describe the slow slip events are still incomplete, which restricts the detailed knowledge of the movements and the associated tremor. 

In the original paper, the strategy adopted by the authors to address the limitation of the current models for the slow slip events was to use Random Forest machine learning algorithm to construct a model capable to predict GPS displacement measurement from the continuous seismic data. This investigation is sustained in the fact that the statistical features of the seismic data are a fingerprint of the fault displacement rate. Therefore, predicting GPS data from seismic data can make GPS measurements a proxy for investigating the fault slip physics and, additionally, correlate this slow slip events with associated tremors that can be studied in laboratory. The purpose of this report is to expose the methodology adopted by the authors and try to reproduce their results as coherent as possible with the original work.




\section{Methodology}

Before starting with the computational exploration of the GPS and seismic data we should keep in mind that this approach is within the 'Big Data' scope, which means that we will have to manipulate and work with large sets of data. Therefore, to get access to all these files it would be helpful to have an automatic way to continuously download a given set of data in spite of download each individual file at time. An efficient manner (when available) to download large sets of data is through the APIs (\textit{Application Programming Interface}). An API is a set of definitions and protocols for building and integrating application software that allows users to get access to contents from servers.

Both seismic data [3] and The GPS data [4] are available through APIs services and, therefore, can be downloaded by following the requesting protocol to the respective domain/address. An efficient way to request data from APIs is with the Pyhton's package \textit{requests}. \textit{requests} can requests different types of encoded data (HTML, XML, \textit{json}, for example) and has some functions that allow users to transform the request response into Python objects.

\subsection{Seismic data download} \label{down}

As previously mentioned the seismic data used in this report is accessible for free download in the IRIS API documentation [3] Following the approach of the original paper, our interest is in the stations near Vancouver Island. We searched for all station within a radius of 0.6\textsuperscript{o} from 48.9\textsuperscript{o} N and -123.9\textsuperscript{o} W. For each station, the East component (HHE channels) raw seismic data were downloaded for every day (00:00 to 01:00) between 01/01/2010 to 30/09/2020. 

\subsection{GPS data download}

The GPS data used in this report are from the Western Canada Deformation Array, preprocessed by the United States Geological Survey. As already mentioned, these data are also accessible through API service [4] and, therefore, can be downloaded using Python \textit{request}. We selected all stations within a square of $\pm$ 1\textsuperscript{o} of 48.9\textsuperscript{o} N, -124.4\textsuperscript{o} W. This GPS data basically represents the set of daily GPS position measurements in a fixed North America reference frame that uses the stable interior reference of the North America continent as a reference [2]. 

\section{Data cleaning and structuring}

Given that both seismic and GPS data are very noisy, before applying the machine learning method we need to adjust and clean the data set. For seismic data, this procedure consists in detrending and demeaning the raw seismic data.

Once the demean and detrend filters were applied, the seismic data was then subdivided into 5 different frequency bands: 8-9 Hz, 9-10 Hz, 10-11 Hz, 11-12 Hz and 12-13 Hz (figure 4). For each
frequency band, four statistical quantities were calculated: skewness, kurtosis, value range (difference between the highest and the lowest amplitude value) and average value. These quantities are stored in matrix form, with the columns formed by the values of the statistical quantities and the lines corresponding to each day of measurement. Given that that six seismic stations were selected according to the criteria previously (section \ref{down}) discussed, the total number of columns in the matrix is equal to 120 (6 stations $\times$ 5 frequency bands $\times$ 4 statistical quantities). In terms of machine learning, these data sets constitute the features of the seismic data. A last stage of data smoothing is the application of rolling mean, which consists of grouping the data in 60-day intervals and calculating the average value of the interval.

Now that we have the features, we must now filter the GPS data, which in terms of machine learning are the target (i.e., values we intend to predict form the seismic features). For each GPS station initially selected, the total data measured by the station was downloaded. Only the data whose dates satisfy the condition of being active between 2010 and 2020 were selected. After this step, the total horizontal displacement (east-west component + north-south component) was calculated for each measurement day. This first data set represents the noisy GPS data. The GPS signal denoise process consists of applying the rolling linear regression: it takes 60 days (as done with the seismic data), then a linear regression is calculated with the values within 60 days range; the slope of the adjustment in each interval defines the displacement rate. These are the data that will be used as a target for the application of the machine learning method.

With the seismic and GPS data, now properly treated and filtered, we must group them all in a matrix whose first 120 contains the values of the statistical quantities of the seismic data (features), and whose 121\textsuperscript{th} line contains the displacement rate values for each GPS station (target). Once the final matrix has been built, containing both the features and the target, the next step is the application of the machine learning method.

\vspace*{0.5cm}
\section{Random Forest implementation}

As highlighted in the original paper, several machine learning algorithms are able to predict output from inputs. For this study, the authors adopted the Random Forest algorithm, which follows the logic of decision trees: from an initial decision, the algorithm sequentially generates new branches according to the stocastically generated criteria. At the end of the process, the final model given by the random forest 'branch' of the tree decision that most closely relates to the target data (i.e., GPS data).

To apply the Random Forest algorithm, we first need to divide the data set (features + label) in two: a part called a training set and another called a validation set. The training set represents the portion of the data in which the Random forest will 'train' to 'learn' how to predict GPS data. For this purpose, the algorithm has access to both GPS data and seismic data (i.e., it 'knows' the results it must achieve). Once 'trained', the effectiveness of the algorithm is put to the test with the data from the validation set. In this step, the algorithm has access only to seismic data (features) and not to GPS data (target).



\vspace*{0.5cm}

\section{Results and conclusions}

Figures 1 to 3 illustrate the final results, where the points represent the estimates obtained by the machine learning algorithm, while the solid red line represents the actual data measured by the GPS stations. The model was generated for each GPS station separately. As we can see, the correspondence between model and real data is surprisingly good. This high correspondence, however, must be associated with some mistake made during the assembly of the final matrix on which the algorithm was applied. A factor that reinforces the possibility of misunderstanding in the application is the fact that even for a single interaction, the model generated is very close to the real values (see appendix for Python code). This mistakes must be correct in order to get valuable information from the seismic data and test the validation of Random Forest algorithm to predict GPS data from noisy seismic data.


\begin{figure}[h!]
     \centering
     \begin{subfigure}[b]{0.48\textwidth}
         \centering
         \includegraphics[width=\textwidth, height = 0.88
         \textwidth]{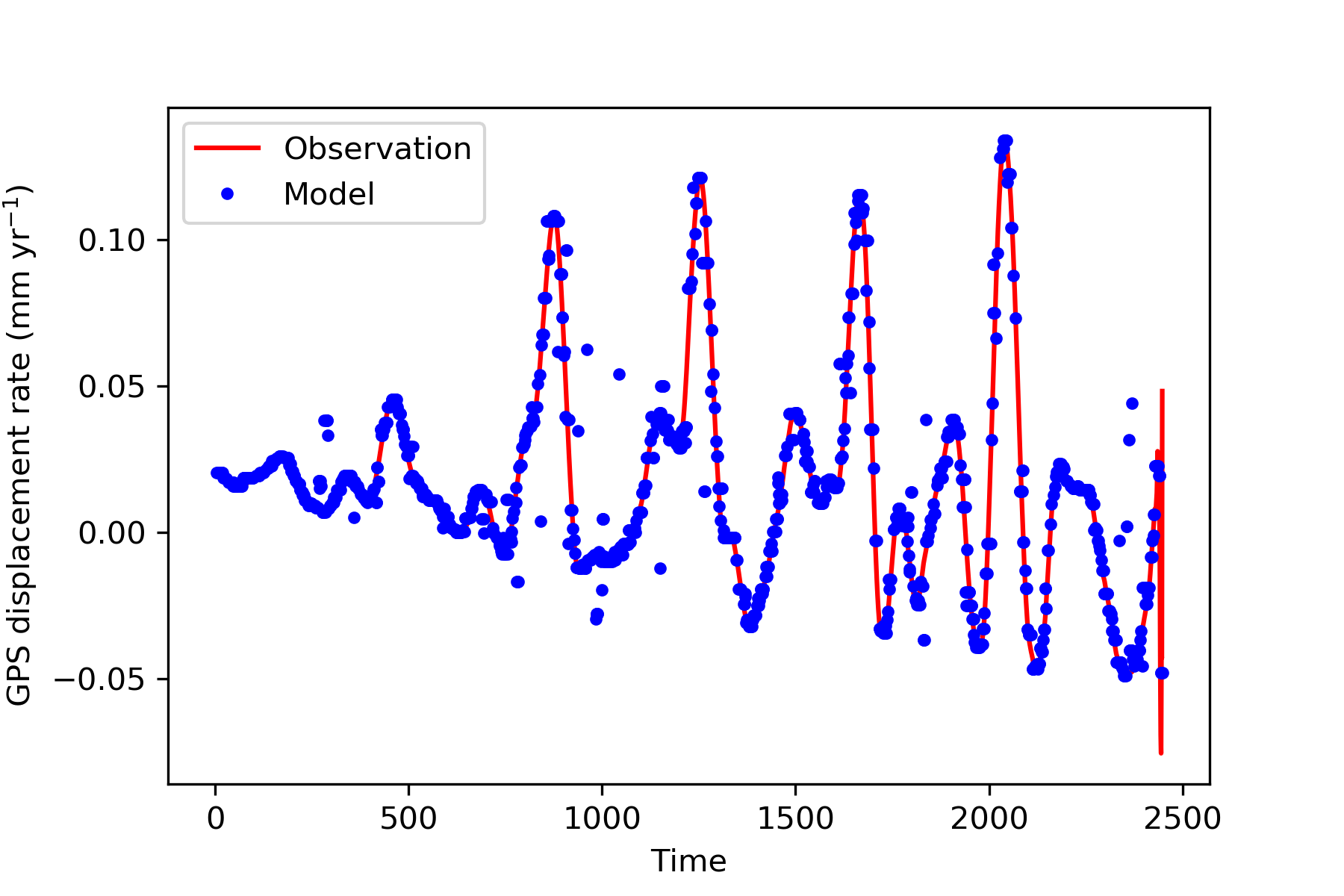}
         \label{fig:dub}
     \end{subfigure}
     \hfill
     \begin{subfigure}[b]{0.48\textwidth}
         \centering
         \includegraphics[width=\textwidth, height = 0.88
         \textwidth ]{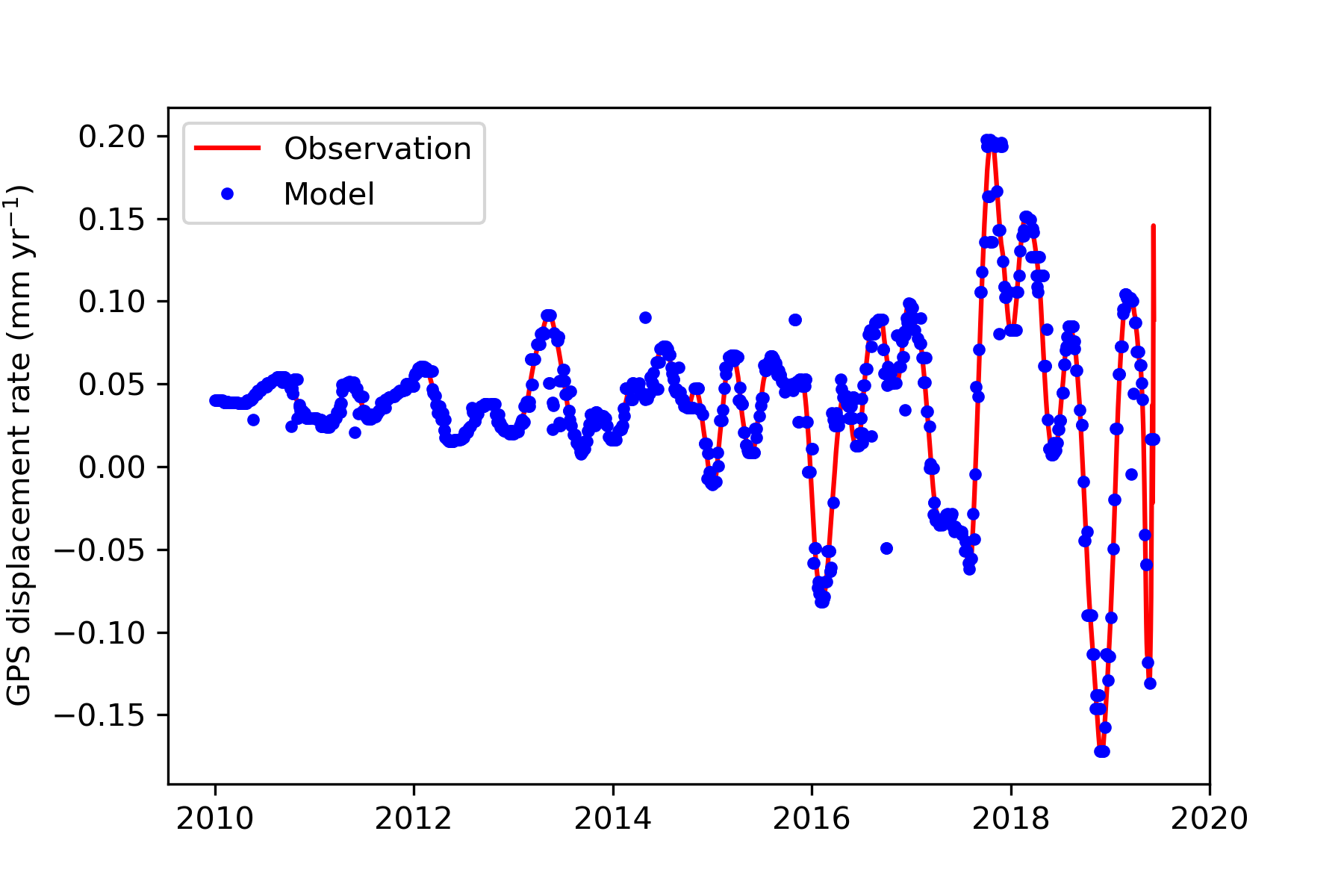}
         \label{fig:1_1}
     \end{subfigure}
     \caption{Modeled $\times$ Real GPS data from Random Forest implementation.}
\end{figure}

\begin{figure}[h!]
     \centering
     \begin{subfigure}[b]{0.48\textwidth}
         \centering
         \includegraphics[width=\textwidth, height = 0.88
         \textwidth ]{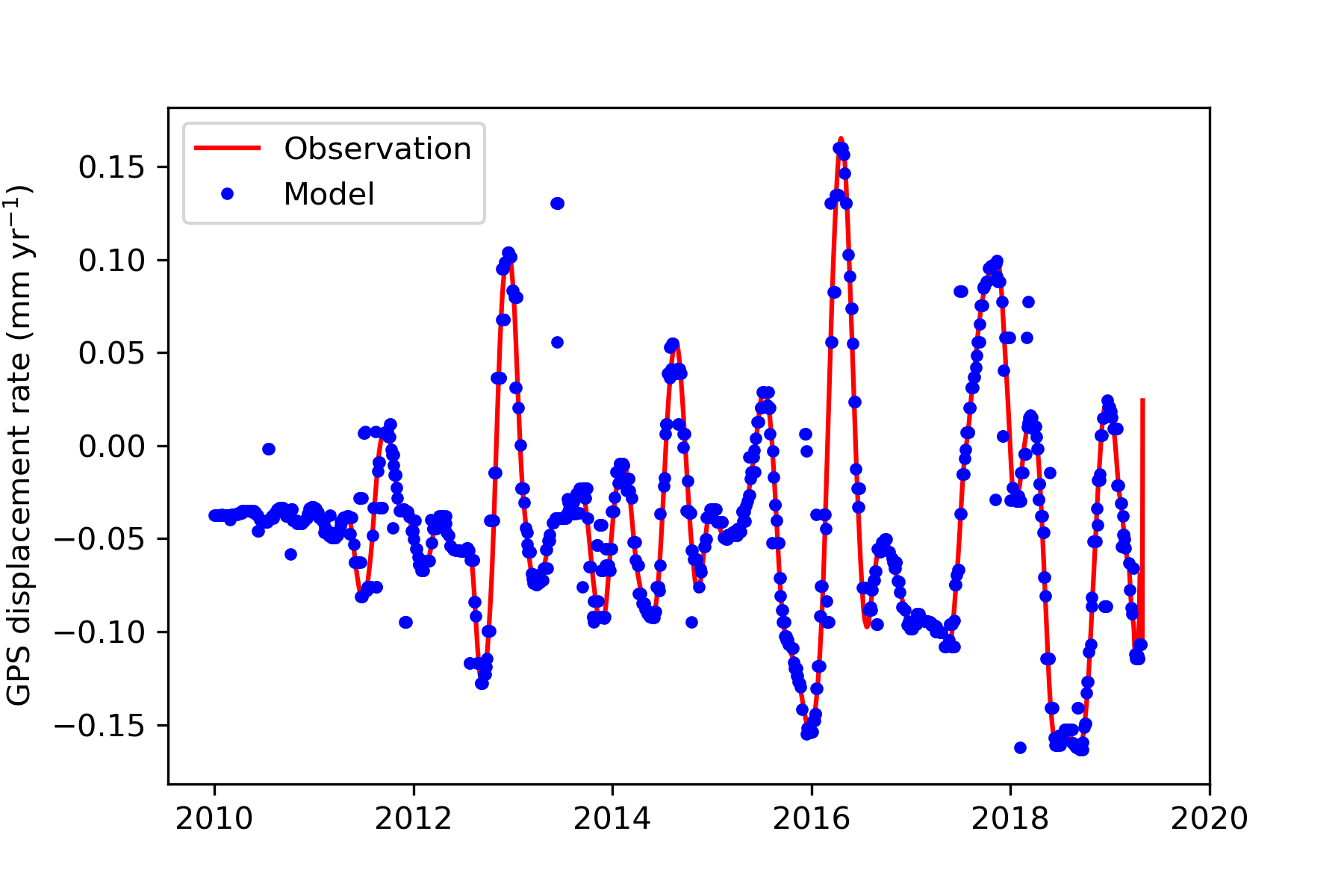}
         
     \end{subfigure}
     \hfill
     \begin{subfigure}[b]{0.48\textwidth}
         \centering
         \includegraphics[width=\textwidth, height = 0.88
         \textwidth ]{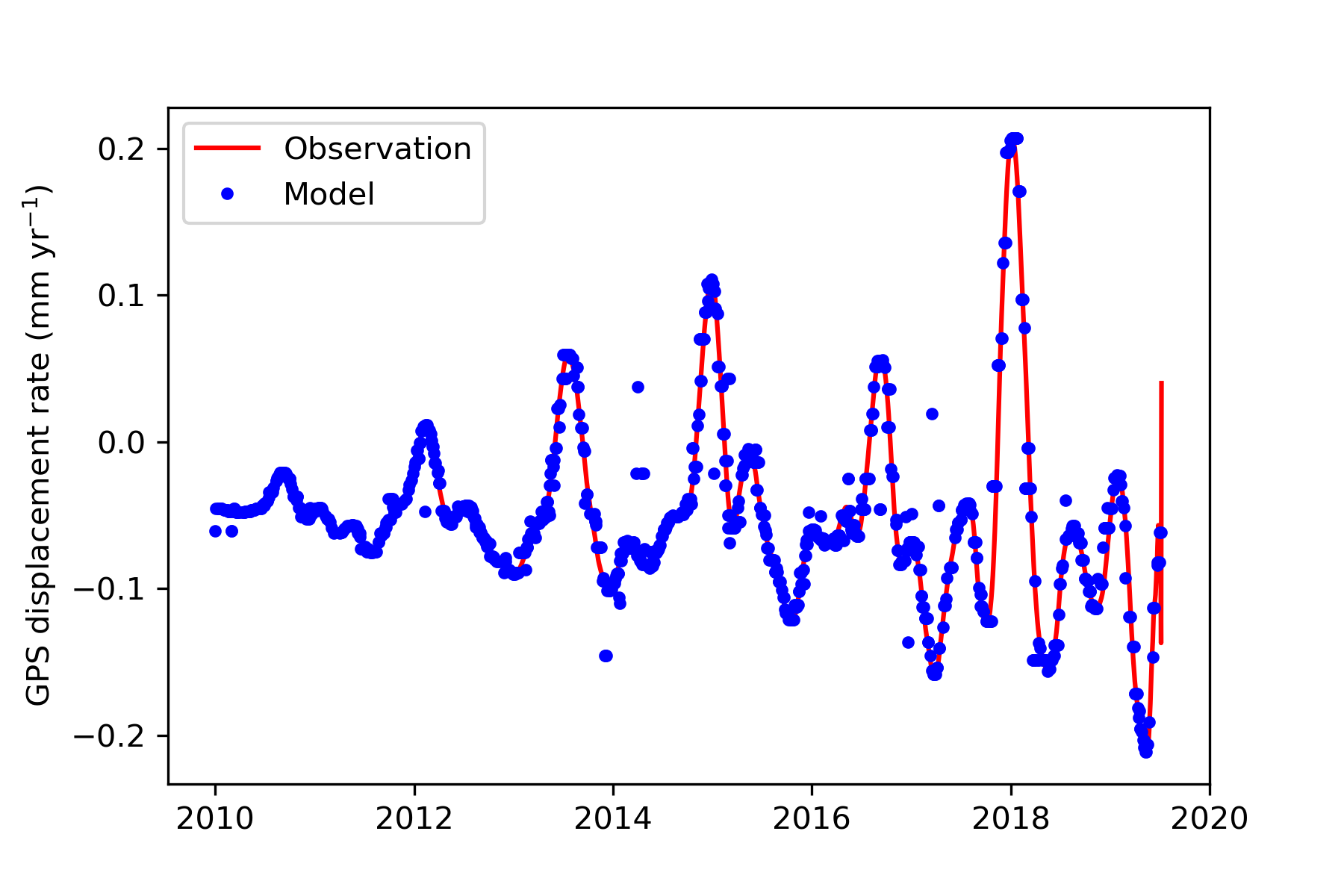}
         \label{fig:2}
     \end{subfigure}
     \caption{Continue.}
\end{figure}

\begin{figure}[h!]
     \centering
     \begin{subfigure}[b]{0.48\textwidth}
         \centering
         \includegraphics[width=\textwidth, height = 0.88
         \textwidth ]{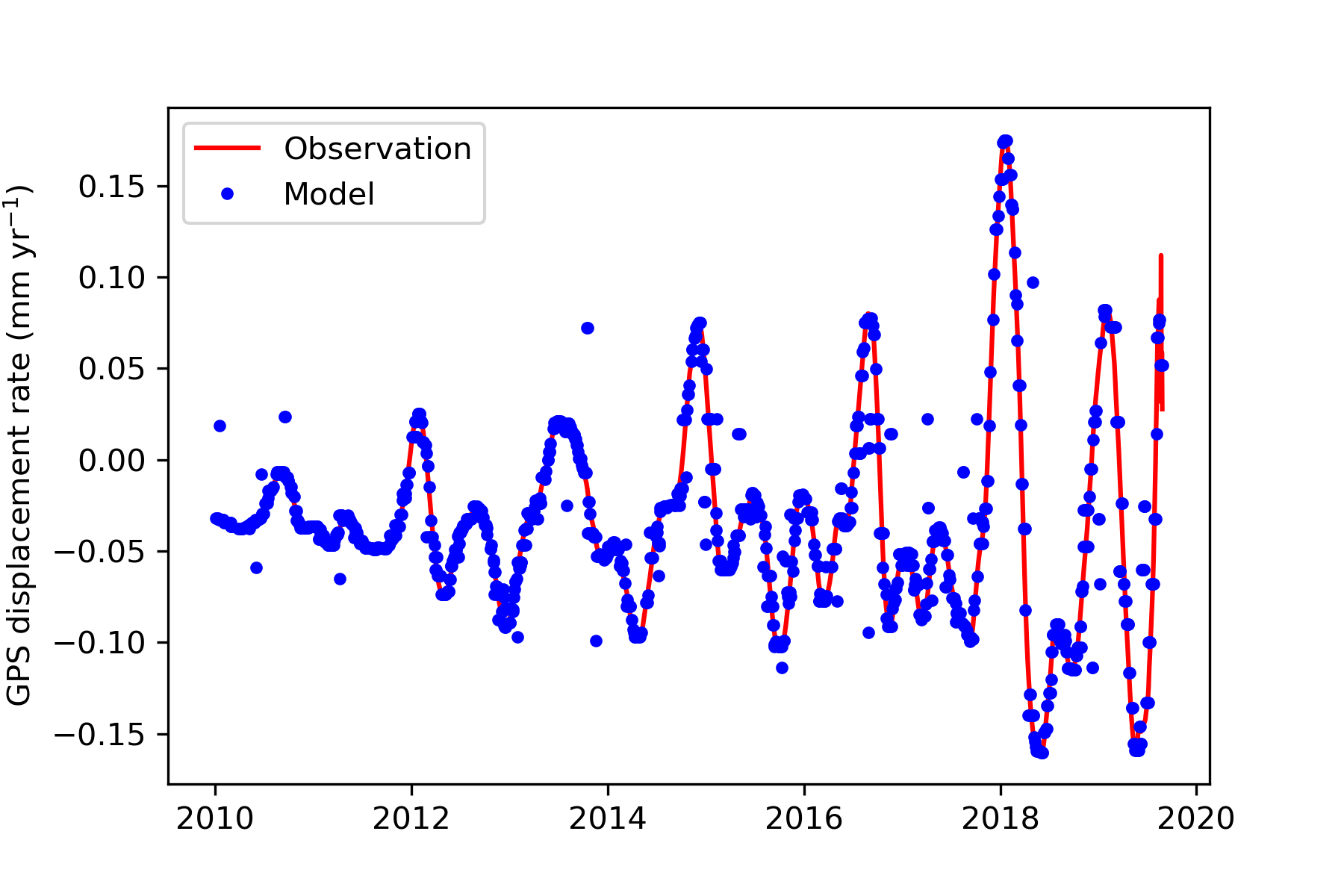}
         
     \end{subfigure}
     \hfill
     \begin{subfigure}[b]{0.48\textwidth}
         \centering
         \includegraphics[width=\textwidth, height = 0.88
         \textwidth ]{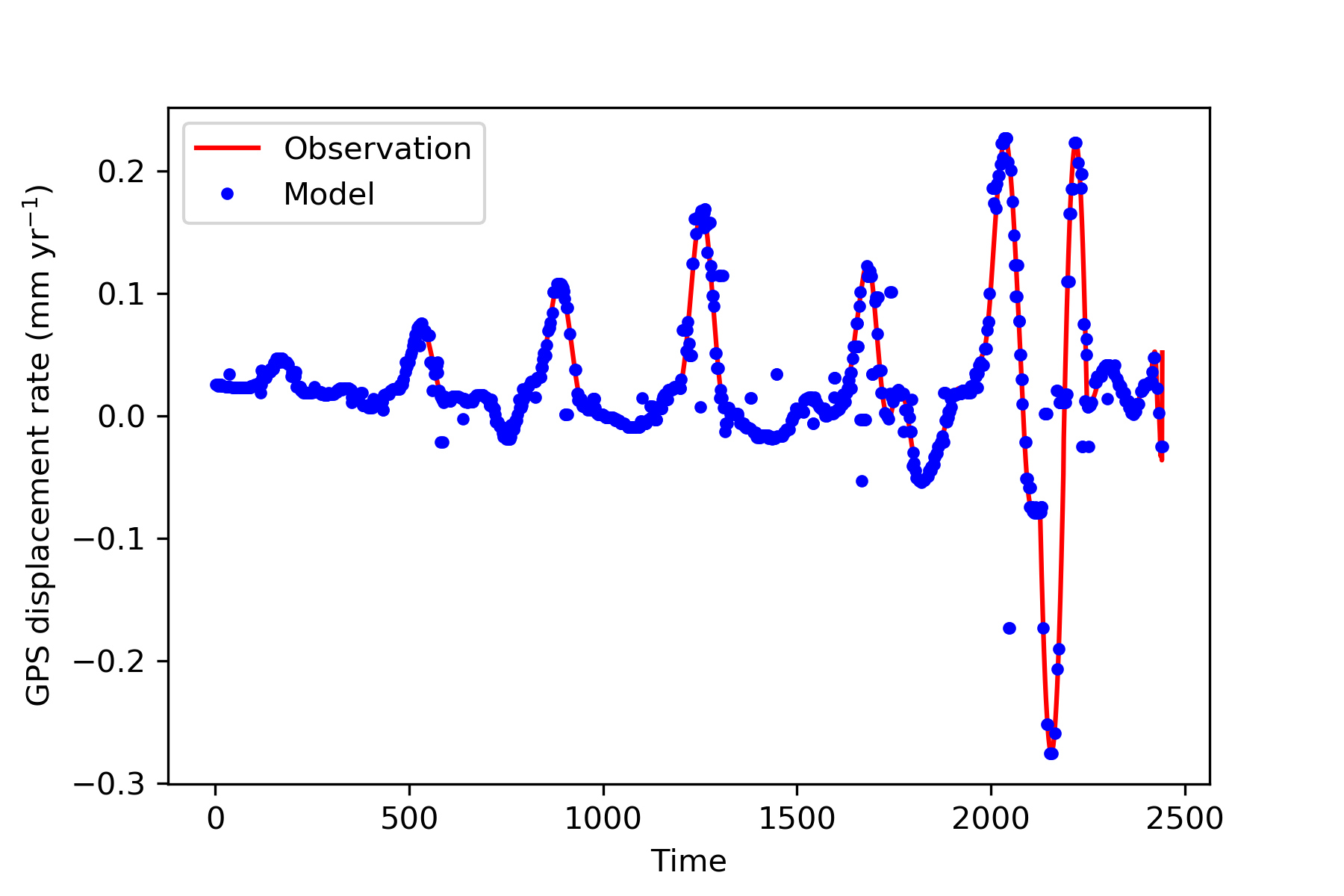}
         \label{fig:3}
     \end{subfigure}
     \caption{Continue.}
\end{figure}

\vspace*{5cm}
\section{References}

[1] Rogers, G. \& Dragert, H. Episodic tremor and slip on the Cascadia subduction zone: the chatter of silent slip. \textit{Science} 300, 1942–1943 (2003).

\vspace{0.1cm}

\noindent [2] Rouet-Leduc, B., Hulbert, C. \& Johnson, P.A. Continuous chatter of the Cascadia subduction zone revealed by machine learning.\textit{Nature Geosci}, 12, 75-79 (2019).

\vspace{0.1cm}

\noindent [3] IRIS DMC FDSNWS Dataselect Web Service; \url{http://service.iris.edu/fdsnws/dataselect/1/}

\vspace{0.1cm}

\noindent [4] Murray, J. R. \& Svarc, J. Global positioning system data collection, processing, and analysis conducted by the US Geological Survey Earthquake Hazards Program. \textit{Seismol. Res. Lett.} 88, 916–925 (2017).

\begin{figure}[h!]
    \centering
    \includegraphics[width=1.\linewidth, height = 1.\linewidth]{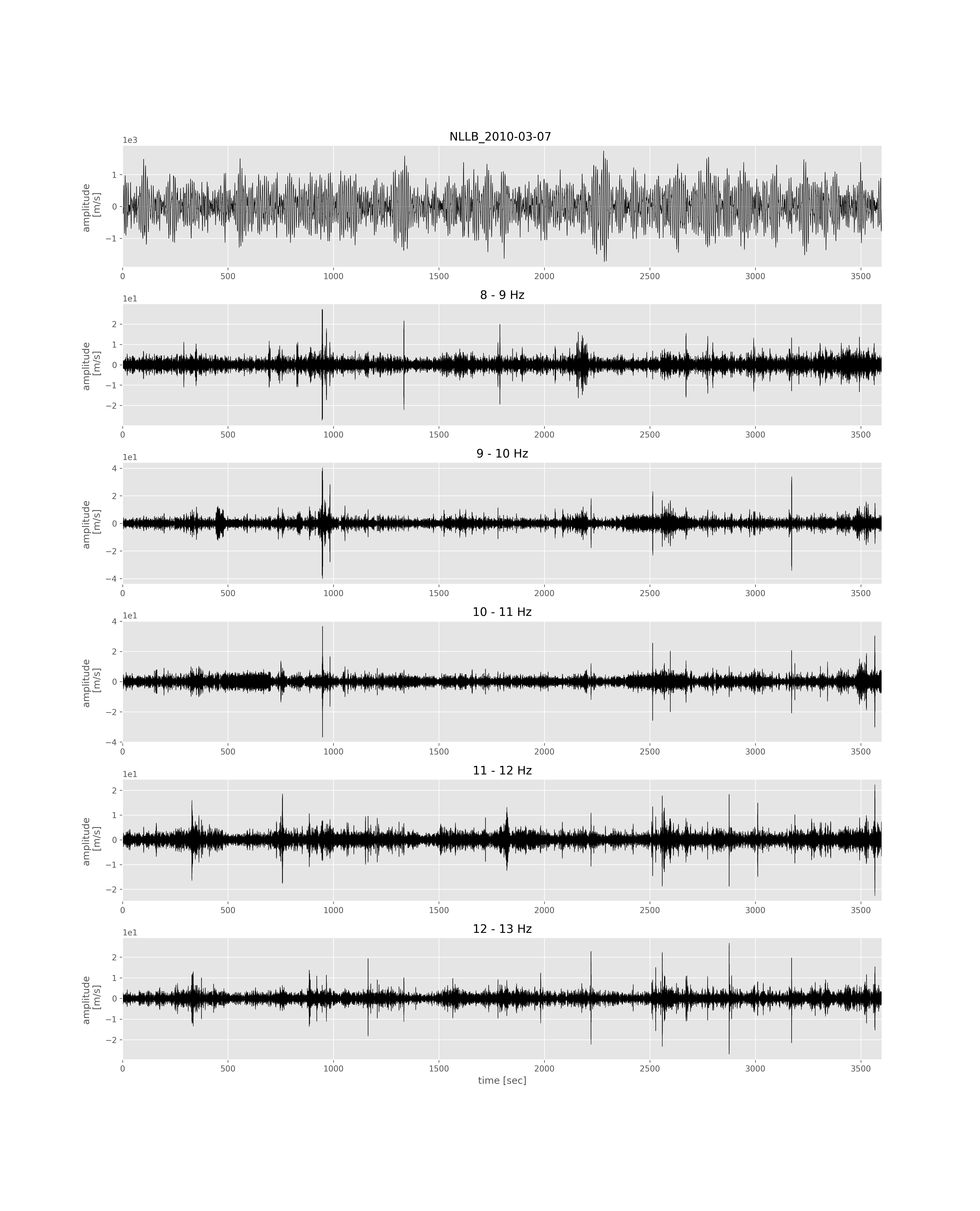}
    \caption{Seismic data from station NLLB in 2010-03-7.}
    \label{fig:seismic}
\end{figure}

\end{document}